\documentclass[conference]{IEEEtran}
\usepackage{mathbbold}
\usepackage{mathrsfs}
\usepackage{amsfonts}
\ifCLASSINFOpdf
  \usepackage[pdftex]{graphicx}
\else
\fi
\usepackage[cmex10]{amsmath}
\usepackage{amssymb}
\usepackage{dsfont}

\begin{document}
%
\title{Diversity and Multiplexing Tradeoff \\in the Uplink of Cellular Systems with Linear MMSE Receiver}

\author{\IEEEauthorblockN{Xiaoshi Song, Changchuan Yin, and Danpu
Liu}
\IEEEauthorblockA{Key Laboratory of Universal Wireless Communications, Ministry of Education,\\Beijing University of Posts and Telecommunications, Beijing, China 100876\\
Email: songxiaoshi@gmail.com, ccyin@ieee.org, dpliu@bupt.edu.cn}}


%



\maketitle

\begin{abstract}
In this paper, we extend the diversity and multiplexing tradeoff (DMT) analysis
from point-to-point channels to cellular systems to evaluate the
impact of inter-cell interference on the system reliability and
efficiency. Fundamental tradeoff among diversity order, multiplexing gain and
inter-cell interference intensity is characterized to reveal the capability of
multiple antennas in cellular systems. And the detrimental effects of the
inter-cell interference on the system performance of diversity and
multiplexing is presented and analyzed.
\end{abstract}
\vspace{10pt}
%
\IEEEpeerreviewmaketitle
\section{Introduction}
Reliability and efficiency are two fundamental aspects of
wireless communication systems. Due to the limitation on the
available channel resources, improvement of either aspect comes at
the price of sacrificing the other. To explore the optimal
reliability when certain system efficiency is required, numerous
channel coding strategies have been studied in single antenna
settings. For multiple-input multiple-output (MIMO) fading channels,
additional spatial dimension is utilized to combat deep fadings for reliability or
increase the available degrees of freedom for spectral efficiency, namely spatial diversity
and multiplexing. It has been proved that multiple
antennas can provide lower error probability and/or higher data rate than conventional single antenna systems \cite{capacity:foschini}-\cite{stbc:Tarokh}. And the tradeoff between the error probability and the data rate can be asymptotically characterized by the tradeoff between the diversity order and the multiplexing gain in the high-power regime.

Fundamental diversity and multiplexing tradeoff is first
characterized in the seminal work of Zheng and Tse \cite{DMT:Zheng},
where an optimal maximum-likelihood (ML) scheme is utilized at the
receiver for data processing. Later, in \cite{DMT MMSE:Kumar}, Kumar
\textit{et al}. consider a low-complexity MIMO architecture and
discuss the DMT performance with suboptimal linear receivers. All
these works focus on the DMT performance in point-to-point settings,
but in practical cellular scenarios, due to the presence of
inter-cell interference, how to characterize the DMT is still an
open problem.

In this paper, we will extend the DMT analysis from point-to-point
channels to cellular systems. Joint spatial
encoding$\!$~\footnote{Although the result still holds in the
separate encoding cases, it is not the subject of this work.} and
linear minimum mean square error (MMSE) MIMO transceiver
architecture \cite{MMSE:A.Hedayat} is utilized for practical
complexity consideration. To evaluate the impact of the inter-cell
interference on the system reliability and efficiency, inter-cell
interference factor $\xi$ is introduced to asymptotically represent the inter-cell interference intensity. And the fundamental tradeoff
among diversity order ${{d}}$, multiplexing gain ${{r}}$ and
inter-cell interference factor $\xi$ is characterized. Given $M$
transmit antennas and $N$ receive antennas (${N \geq M}$), our main
result shows that the optimal DMT in cellular systems is
${d}_{{mmse}}^*\left( {r}, {\xi} \right) = \left(N - M +
1\right)\left({1} - \xi - \frac{r}{M}\right)^{+}$, where
${\left(x\right)^{+}}$ denotes $\max(0,x)$.

The rest of the paper is organized as follows. The system model and
definitions are described in
section~\uppercase\expandafter{\romannumeral 2}. Our main result on
the optimal tradeoff among diversity order, multiplexing gain and inter-cell
interference intensity is derived in
section~\uppercase\expandafter{\romannumeral 3}. In
section~\uppercase\expandafter{\romannumeral 4}, numerical results and interpretations are presented to provide insights on the impact
of  inter-cell interference in cellular systems. And
section~\uppercase\expandafter{\romannumeral 5} summarizes our
conclusions.
\section{System Model}
Consider the uplink of a narrow-band\footnote{For wide-band systems, we can decompose the frequency-selective channel into multiple parallel and independent frequency-flat sub-channels. And, on each sub-channel, the narrow-band assumption is valid.} cellular system, with each user
having $M$ transmit antennas and each base station having $N$ $(N
\geq M)$ receive antennas. At each time instance, only one user is scheduled and served in each cell. And simultaneous transmissions
in different cells cause the inter-cell interference. It is assumed
that there is no cooperation among the cells. For each receiver in
the system, only the local-cell signal is concerned and the
inter-cell interference is simply treated as noise.
We further assume that the
interference at a given base station is caused by the active links
from its adjacent cells (first-tier cells) only.
Suppose each cell
has $K - 1$ adjacent cells. Without loss of generality, we focus on
one certain cell (so called the reference cell) in the system and
assume that the user in the cell communicates with its intended base
station under the presence of $K - 1$ co-channel interferers.

Let us index the desired user in the reference cell by $K$ and index
the interfering users in the adjacent cells by the numbers starting
from $1$ to $K - 1$. At time \textit{t}, given the transmitted
signal vector ${{\bf{x}}_k} \in {\mathbb{C}^{M \times 1}}$ of the
$k$-th user, the received signal vector ${{\bf{y}}} \in
{\mathbb{C}^{N \times 1}}$ at the intended base station can be
written as
\begin{equation}
\label{receive-model}\setcounter{equation}{1} {{\bf{y}}} = \sqrt
{\frac{{{{\sf{SNR}}}}}{{M}}} {{\bf{H}}_K}{{\bf{x}}_K} +
\sum\limits_{\scriptstyle k = 1}^{K - 1}{\sqrt
{\frac{{{{\sf{INR}}_{(k)}}}}{{M}}} {{\bf{H}}_{k}}{{\bf{x}}_k} +
{\bf{n}}},
\end{equation}
where ${\bf{n}} \in {\mathbb{C}^{N \times 1}}$ denotes the complex
Gaussian noise vector with entries $\sim \mathcal {C}\mathcal
{N}({0,1})$, ${{\sf{SNR}}}$ denotes the average signal-to-noise
ratio at each receive antenna of the base station and
${{\sf{INR}}_{(k)}}$ denotes the average interference-to-noise ratio
at each receive antenna of the base station from the ${k}$-th (${k}
= 1,~2,~...,~K - 1$) interfering user.
${\bf{H}}_{k}$ denotes the
frequency-flat channel fading matrix between the ${k}$-th user and
the base station. Under rich scattering condition,
${\bf{H}}_{k}$ is assumed to have independent and identically
distributed entries with zero mean and unit variance. $T$-block
length spatial-temporal codeword ${\bf{X}}_{k}\in\mathcal {C}^{M
\times T}$ is chosen uniformly at random from the common codebook
$\mathcal {C}$ and launched at the ${k}$-th user.
An overall power constraint is considered on $\mathcal {C}$ as
\begin{equation} \label{power constraint}
\frac{1}{{\left| C \right|}}\sum\limits_{j = 1}^{\left| C \right|}
{\left\| {\bf{X}}{(j)} \right\|_F^2}  \le MT,
\end{equation}
where $\left| C \right|$ denotes the size of the codebook,
${\bf{X}}{(j)}$ denotes the $j$-th codeword and $\left\| {{\rm{ }}
\cdot {\rm{ }}} \right\|_F$ denotes the Frobenius norm of a
matrix. For each transmission block, data stream is first encoded, then interleaved and finally multiplexed into $M$
sub-streams. Thus, vector ${{\bf{x}}_{k}} $ can be regarded as the
spatially encoded sample of ${\bf{X}}_{k}$ at time instant ${t}$.

In this paper, quasi-static fading environment is considered that each channel matrix is randomly chosen at the beginning but remains constant over the whole block length ${T}$. And we assume that the channel state information (CSI) between the ${k}$-th user ($k = 1, 2, ..., K $) and the
receiver is perfectly known at the receiver but not available at any
of the users. Though, according to \cite{Capacity:Angel Lozano}, generally, the interference channel matrix ${\bf{H}}_k$, $k = 1, 2, ..., K - 1$, cannot be estimated in cellular systems. In this work, for theoretical analysis, we assume that all the users may periodically send the orthogonalized pilot signals and thus the receivers can obtain the channel matrices accurately.
According to \cite{capacity:telatar} and
\cite{DMT:Zheng}, without loss of optimality, we assume the distribution of
${\bf{x}}_{k}$ to be circularly symmetric complex Gaussian
with zero mean and covariance matrix ${\bf{I}}$, where $\bf{I}$ is
the identity matrix.

We aim to characterize the fundamental tradeoff among the diversity order, multiplexing gain and inter-cell interference intensity. According to \cite{DMT:Zheng}, we introduce the definition of  \textit{multiplexing gain} ${r}$
and \textit{diversity gain} ${d}$ as follows:

\textit{Definition 1:} A scheme ${\mathcal {C}(\sf{SNR})}$ operating at $\sf{SNR}$ is said
to achieve spatial multiplexing gain ${r}$ and diversity gain ${d}$
if the data rate ${{\mathcal
{R}}(\sf{SNR})}$ satisfies
\begin{equation}
\label{multiplexing gain} \mathop {\lim
}\limits_{\sf{SNR}\rightarrow \infty} \frac{{{\mathcal
{R}}(\sf{SNR})}}{{\log \sf{SNR}}} = {r}
\end{equation}
and the average error probability ${\mathcal {P}_e}(\sf{SNR})$ satisfies
\begin{equation}
\label{diversity gain} \mathop {\lim }\limits_{\sf{SNR} \rightarrow
\infty} \frac{{\log {\mathcal {P}_e}(\sf{SNR})}}{{\log \sf{SNR}}} =
- {d}.
\end{equation}
Similarly, according to \cite{Interference:Etkin}, we introduce the inter-cell interference factor ${\xi}_{(k)}$ to asymptotically measure the inter-cell interference intensity of the $k$-th interferer ${\sf{INR}}_{(k)}$ with respect to $\sf{SNR}$ as
\begin{equation}
\label{interference factor}\mathop {\lim }\limits_{{\sf{SNR}} {\to}
\infty } \frac{{{{\log }}{{}}{{{\sf{INR}}_{(k)}}}}}{{{{\log }}{{\sf{SNR}}}}}
= {\xi}_{(k)},
\end{equation}
where
\[0 < \xi_{(k)} < 1.\]
We further introduce the notation of exponential equality $f({\sf{SNR}}) \buildrel\textstyle.\over= {\sf{SNR}}^b$ to denote
\[
\mathop {\lim
}\limits_{\sf{SNR}\rightarrow \infty} \frac{{f(\sf{SNR})}}{{\log \sf{SNR}}} = {b}.
\]
And we say that $f({\sf{SNR}})$ is exponential equal to $g({\sf{SNR}})$, i.e. $f({\sf{SNR}}) \buildrel\textstyle.\over= g({\sf{SNR}})$, if
\[
\mathop {\lim
}\limits_{\sf{SNR}\rightarrow \infty} \frac{{f(\sf{SNR})}}{{\log \sf{SNR}}} = \mathop {\lim
}\limits_{\sf{SNR}\rightarrow \infty} \frac{{g(\sf{SNR})}}{{\log \sf{SNR}}}.
\]
Accordingly, the symbols $\buildrel\textstyle.\over\le$ and $\buildrel\textstyle.\over\ge$ are defined.

In this work, we assume that the inter-cell interference ${\sf{INR}}_{(k)}$, for $k = 1, 2, ..., K-1$, are exponentially equivalent that
\[{{\sf{INR}}_{(1)}} \buildrel\textstyle.\over= {{\sf{INR}}_{(2)}} \buildrel\textstyle.\over= ... \buildrel\textstyle.\over= {{\sf{INR}}_{(K - 1)}} \buildrel\textstyle.\over= \sf{INR} \buildrel\textstyle.\over= {\sf{SNR}}^\xi.\]
It can be interpreted as: though the interference caused by the co-channel interferers may differ from each other, they have the same rate of change $\xi$ with respect to the SNR of the desired link. Then, for asymptotic analysis, we rewrite (\ref{receive-model}) as
\begin{equation}
\label{asympotic receive-model}{{\bf{y}}} = \sqrt
{\frac{{{{\sf{SNR}}}}}{{M}}} {{\bf{H}}_K}{{\bf{x}}_K} + \sqrt
{\frac{{{{\sf{INR}}}}}{{M}}} \sum\limits_{\scriptstyle k = 1}^{K - 1}{\sqrt
{\frac{{{{\sf{INR}}_{(k)}}}}{{\sf{INR}}}} {{\bf{H}}_{k}}{{\bf{x}}_k} +
{\bf{n}}}.
\end{equation}

Since for good codes and long block length \textit{T}, the codeword error
event can be characterized by the channel outage event at high
{\rm{SNR}} region, we use the outage probability $\Pr_{out}$ instead of error probability ${\mathcal {P}_e}$ in (\ref{diversity gain}) to capture the diversity behavior. The outage probability is defined as the probability that the mutual information of the channel fails to support the target rate. To be specific, given the mutual information ${\mathcal
{I}}$ and the target rate ${\mathcal
{R}}$, we have the outage probability $\Pr_{out}$ as
\begin{equation} \label{outage Probability}
{\Pr}_{out}({\mathcal
{R}}) = \Pr \left( {\mathcal
{I}} \le {\mathcal
{R}}\right).
\end{equation}

In \cite{DMT:Zheng}, with ML detector, the
diversity-multiplexing tradeoff function ${{d}^*}\left( {r} \right)$
in a general point-to-point scenario is derived as
\[{{d}^*}\left( {r} \right) = (M - {r})(N - {r}),\]
where \[{r} = 0,~1,~...,~\min\{M,N\}.\]
In this work, instead of using ML receiver, we consider the
suboptimal linear MMSE receiver and derive the optimal DMT. The MMSE
receiver is characterized by a matrix ${\bf{G}}$ which minimizes the
mean square error between the transmit signal ${\bf{x}}$ and its estimation ${{\hat
{\bf{x}}}} = {\bf{G}\bf{y}}$ as
 \[{\bf{G}} = \arg{\mathop {\min }\limits_{{\bar{\bf{G}}}}} E\left[ {\left\| {{\hat {\bf{x}}}} - {\bar{\bf{G}}\bf{y}} \right\|}^{2} \right]
.\]
Using the orthogonality principle, we can then have $\bf{G}$ as
\begin{equation}
\label{mmse receiver} {\bf{G}} = \sqrt{\frac{M}{\sf{SNR}}}{{\bf{H}}^H}{\left( {{\bf{HH}}_{}^H
+ \frac{{\sf{INR}}}{{\sf{SNR}}}{\bf{R}}{\bf{Q}}{{{\bf{R}}}^H} +
\frac{{M}}{{\sf{SNR}}}{{\bf{I}}}} \right)^{ - 1}},
\end{equation}
where ${\left( {{\rm{ }} \cdot {\rm{ }}} \right)^H}$ denotes the
conjugate transpose, ${\left( {{\rm{ }} \cdot {\rm{ }}} \right)^{ -
1}}$ denotes the matrix inversion, ${\bf{H}} = {{\bf{H}}_K}$, ${\bf{Q}} = {\rm{blockdiag}}{\{\frac{{\sf{INR}}_{(k)}}{\sf{INR}}{\bf{I}}\}}$ and ${\bf{R}} = \left[ {{{\bf{H}}_1},{{\bf{H}}_2},...,{{\bf{H}}_{K-1}}}\right]$.

Actually,  by applying ${\bf{G}}$, the channel matrix of the transmit-receive
pair is divided into ${M}$ parallel pipes, which is alternatively
called virtual channels. For each virtual channel, one complex
degree of freedom is supported. And, by using Woodbury's identity
\cite{MMSE:M.R.McKay}, the $\rm{SINR}$ of the ${i}$-th channel $\rho_i$ is
shown to be
\begin{equation}
\label{MMSE SINR} \!\!{{\rho}}_i = \left(\!{{{{\left[
{{{\left(\!{{{\bf{I}}}\!\! + \!\!{\frac{
{\sf{SNR}}}{\sf{INR}}{{\bf{H}}^H}\!{{\left(\!{\bf{R}}{\bf{Q}}{{{\bf{R}}}^H} \!\!+\!
\frac{{M}}{{\sf{INR}}}{{\bf{I}}} \right)\!\!}^{ - 1}}\!\!{\bf{H}}}}
\!\right)}^{ - 1}}} \right]}_{i,i}}}}\right)^{\!-1} \!\!\!\!\!\!- 1,
\end{equation}
where ${\left[ {{\rm{ }} \bf{A} {\rm{ }}} \right]_{i,i}}$ denotes the
${i}$-th diagonal entry of matrix $\bf{A}$.
\section{Diversity-Multiplexing Tradeoff}
In this section, we extend the DMT results in \cite{DMT MMSE:Kumar}
from point-to-point channels to cellular systems to evaluate the
impact of the interference on system reliability and efficiency. And our
main result is given as follows:\

 \textit{Theorem 1:} Consider the uplink of a cellular system, at a given time
${t}$, the user in a certain cell communicates with its intended
base station
under the presence of the co-channel interference from its
$K - 1$ adjacent cells.
And the optimum
diversity-multiplexing tradeoff for the transmit-receive pair in the
cell is given by:
\begin{equation}
\label{MMSE DMT} {d}_{{mmse}}^*\left( {r}, \xi \right) = \left(N - M
+ 1\right)\left({1} - \xi - \frac{{r}}{M}\right)^{+},
\end{equation}
where joint spatial Gaussian encoding is applied across the transmit
antennas and linear MMSE equalizer is utilized at the receiver.

\textit{Proof :} We first give an upper and a lower bound on the
outage exponent, then prove the theorem by using the squeeze lemma.

\textbf{Lower bound on the outage exponent.} Since joint spatial
encoding scheme is utilized, data transmission is in outage only
when the aggregate mutual information fails to support the target
data rate ${{\mathcal {R}}}$. Given ${\bf{H}}$ and ${{\bf{R}}}$, the
mutual information of the transmit-receive pair in the cell by using
linear MMSE receiver is given by
\begin{equation}
\label{mmse mutual information}
\begin{split}
{{\bf{I}}_{mmse}}\left(
{{\bf{H}},{\bf{R}}} \right) &= - \sum\limits_{i = 1}^M {\log \left( 1 + \rho_i \right)}\\
& = - \sum\limits_{i = 1}^M { \log \left( {{{\left[
{{{\left( {{\bf{I}} +
\!\frac{{\sf{SNR}}}{{{}\sf{INR}}}}\bf{C} \right)}^{ - 1}}}
\right]}_{i,i}}}\right)},
\end{split}
\end{equation}
where
\begin{equation}
\label{Quadradic Form}
{\bf{C}} = {{\bf{H}}^H}{{\left(
{{\bf{R}}{\bf{Q}}{{{\bf{R}}}^H} +
\!\frac{{M}}{{{}\sf{INR}}}{\bf{I}}} \right)}^{ -
1}}{\bf{H}}.
\end{equation}
Since the function $-\log(\cdot)$ is convex, applying Jensen's inequality, (\ref{mmse mutual
information}) can be written as
\[\!\!{{\bf{I}}_{mmse}}\left(
{{\bf{H}},{\bf{R}}} \right) \ge - {M} {\log \left( \frac{1}{{M}} \sum\limits_{i = 1}^M{{{\left[ {{{\left( {{\bf{I}} + \frac{{\sf{SNR}}}{{{}\sf{INR}}}{\bf{C}}} \right)}^{- 1}}}
\right]}_{i,i}}}\right)}\]
\begin{equation}
\label{mmse mutual information at high SNR}
\quad\qquad\quad \;\;\:\,\;\:\:\,\, = - {M} {\log \left(
\frac{1}{{M}}{{\rm{Tr}}}{{{\left[ {{{\left( {{\bf{I}}} +
{\frac{{\sf{SNR}}}{{{}\sf{INR}}}{\bf{C}}} \right)}^{- 1}}}
\right]}}}\right)}.
\end{equation}
Let ${{\lambda _1} \ge {\lambda _2} \ge ... \ge {\lambda _M}}$
denote the ordered eigenvalues of $\bf{C}$,
and (\ref{mmse mutual information at high SNR}) reduces to
\[{{\bf{I}}_{mmse}}\left(
{{\bf{H}},{\bf{R}}} \right) =  - M\log \left(
{ \frac{1}{M}\sum\limits_{i =
1}^M {\frac{1}{{{1 + \frac{\sf{SNR}}{\sf{INR}} \lambda _i}}}} } \right)\]
\begin{equation}
\label{mmse mutual information with eigen value}
\quad\quad\quad\quad\ge - M\log \left({
\frac{1}{{{1 + \frac{{\sf{SNR}}}{{\sf{INR}}}\lambda_M}}}} \right).
\end{equation}
Then, the outage probability is given as
\[{\Pr} _{out}^{mmse}({\mathcal
{R}}) = \Pr \left( {{{\bf{I}}_{mmse}}({\bf{H}},{\bf{R}}) \le
{\mathcal {R}}} \right) \qquad \qquad \qquad \qquad
\]
\[\quad \quad\,\,\,\, \le \Pr \left( { - M\log \left({
\frac{1}{{{1 + \frac{{\sf{SNR}}}{{\sf{INR}}}\lambda _M}}}} \right) \le {\mathcal {R}}} \right)\]
\[{ \qquad \quad \,\;\,\,\,\,\,\,\, } = \Pr \left( {\log \left({
\frac{1}{{{1 + \frac{{\sf{SNR}}}{{\sf{INR}}} \lambda _M}}}} \right) \ge  -
\frac{{{r\log{\sf{SNR}}}}}{{M}}} \right)\]
\begin{equation}
\label{outage probability} \qquad\qquad{\,} = \Pr \left(
\frac{1}{{{1 + \frac{{\sf{SNR}}}{{\sf{INR}}}\lambda _M}}} \ge {{\sf{SNR}}^{-\frac{r}{M}}} \right). { \quad \quad }
\end{equation}
And, we can obtain the asymptotic upper bound on the outage probability as
\[
{\Pr} _{out}^{mmse}({\mathcal
{R}}) \buildrel\textstyle.\over\le
\Pr \left( \frac{{\sf{INR}}}{{\sf{SNR}}} \cdot \frac{1}{{{\lambda _M}}} \ge {{\sf{SNR}}^{-\frac{r}{M}}} \right)
\]
\begin{equation}
\label{asymptotic outage probability upper bound}
\qquad \quad \,\:\,\buildrel\textstyle.\over= \Pr \left( {\lambda _M} \le {{\sf{SNR}}^{\frac{r}{M} + \xi - 1}} \right).
\end{equation}
It should be noticed that the above asymptotic outage probability upper bound vanishes to zero as \rm{SNR} goes to infinity only when ${r \mathord{\left/
 {\vphantom {1 2}} \right.
 \kern-\nulldelimiterspace} M} < 1- \xi$. For ${r \mathord{\left/
 {\vphantom {1 2}} \right.
 \kern-\nulldelimiterspace} M} \ge 1- \xi$, the asymptotic outage probability upper bound approaches 1 and thus the outage exponent lower bound is zero.
When ${r \mathord{\left/
 {\vphantom {1 2}} \right.
 \kern-\nulldelimiterspace} M} < 1- \xi$, we have
\[
\Pr \left( {\lambda _M} \le {{\sf{SNR}}^{\frac{r}{M} + \xi - 1}} \right) = \int_0^{{\sf{SNR}}^{\frac{{r}}{M} + \xi - 1}} {{f_{{\lambda _M}}}\left( {{\lambda _M}} \right)d{\lambda _M}}.
\]
According to \cite{Capacity:Liang Sun}, the first-order expansion of the marginal probability density function (pdf) of $\lambda_M$ is
given by \footnote{We say that $f(x) = o(g(x))$ if ${f(x) \mathord{\left/
 {\vphantom {1 2}} \right.
 \kern-\nulldelimiterspace} g(x)} \rightarrow 0$ as $x \rightarrow 0$ \cite[eq.~1.3.1]{Small o:N. G. Bruijn}.}
\begin{equation}\label{pdf of lambda}
{f_{{\lambda _M}}}\left( {{\lambda _M}} \right) = {a_M}\left( {N - M
+ 1} \right)\lambda _M^{N - M} + o\left( {\lambda _M^{N - M + 1}}
\right),
\end{equation}
where ${a_M}$ is the normalizing factor which is independent of
$\lambda _M$ and $\rm{SNR}$. As such, for $\lambda \ll 1$, (\ref{asymptotic outage probability upper bound}) can
be further written as
\[{\Pr} _{out}^{mmse}(R) \le \int_0^{{\sf{SNR}}^{\frac{{r}}{M} + \xi - 1}} {{f_{{\lambda _M}}}\left( {{\lambda _M}} \right)d{\lambda _M}} \]
\begin{equation}\label{solution of pout}
 \quad\quad\quad\quad\quad\!\!= {a_M}{\sf{SNR}}^{\left( {N - M + 1} \right)\left( {\frac{{r}}{M} +
 \xi - 1} \right)}.
\end{equation}
Therefore, we obtain a lower bound on the outage exponent as
\begin{equation}\label{outage exponent lower bound}
{{d}_{mmse}^{*}}\left( {r}, \xi \right) \ge \left( {N - M + 1}
\right)\left( {1 - \xi - \frac{{r}}{M}} \right)^{+}.
\end{equation}

\textbf{Upper bound on the outage exponent.}
Due to the concavity of $\log( \cdot )$ function, by applying Jensen's inequality on (\ref{mmse mutual information}), we can obtain
\begin{equation}\label{mmse mutual information upper bound}
\!\!\!\!{{\bf{I}}_{mmse}}\left(
{{\bf{H}},{\bf{R}}} \right) \le M \!\log \!\left( {\frac{1}{M}\sum\limits_{i =
1}^M {\frac{1}{{{{\left[ {{{\left( {{\bf{I}} +
\frac{\sf{SNR}}{\sf{INR}}{\bf{C}}} \right)}^{ - 1}}}
\right]}_{i,i}}}}} } \!\right)\!.
\end{equation}
Since ${\bf{C}}$ is a hermitian matrix, we consider the decomposition ${\bf{C}} = {{\bf{U}}^{H}{\bf{\Lambda}}{\bf{U}}}$ on ${\bf{C}}$,
where ${\bf{U}} \in {\mathbb{C}^{ M \times M }}$ is unitary
and ${\bf{\Lambda}} \in {\mathbb{C}^{ M \times M }}$ denotes
a diagonal matrix with eigenvalues of $\bf{C}$ on the diagonal.
Let ${{\bf{u}}_{i}}$ denotes the \textit{i}-th column of {\bf{U}}, then
\[{{{{\left[ {{{\left( {{\bf{I}} +
\frac{\sf{SNR}}{\sf{INR}}{\bf{C}}} \right)}^{ - 1}}}
\right]}_{i,i}}}} = {{\bf{u}}_i^H}\left({\bf{I} + \frac{\sf{SNR}}{\sf{INR}}{\bf{\Lambda}}}\right)^{-1}{{\bf{u}}_i}\]
\[ \;\qquad\qquad\quad\quad= \sum\limits_{j = 1}^M {\frac{{{{|{{u_{j,i}}}|}^2}}}{{1 + \frac{\sf{SNR}}{\sf{INR}}{\lambda _j}}}}. \]
And, according to \cite{DMT MMSE:Kumar}, we have
\[\!\!\!\!\!\!\!\!\!\!\!\!\!\!\!\!\!
{\frac{1}{M}\sum\limits_{i =
1}^M {\frac{1}{{{{\left[ {{{\left( {{\bf{I}} +
\frac{\sf{SNR}}{\sf{INR}}{\bf{C}}} \right)}^{ - 1}}}
\right]}_{i,i}}}}}} = \frac{1}{M}\sum\limits_{i =
1}^M {\frac{1}{\sum\limits_{j = 1}^M {\frac{{{{|{{u_{j,i}}}|}^2}}}{{1 + \frac{\sf{SNR}}{\sf{INR}}{\lambda _j}}}}}}
\]
\[
\qquad\qquad\qquad\quad\qquad\qquad\le \frac{1 + \frac{\sf{SNR}}{\sf{INR}}\lambda_M}{M}\sum\limits_{i =
1}^M\frac{1}{{|{{u_{M,i}}}|}^2}.
\]
Let ${{\mathcal {A}}}$ denote the event
\{$\frac{1}{M}\sum\limits_{i =
1}^M\frac{1}{{|{{u_{M,i}}}|}^2} \le c$\}, where ${c}$ is some constant independent of ${\sf{SNR}}$.
And it has been proved in \cite{DMT MMSE:Kumar} that $\Pr \left({{\mathcal {A}}} \right)$ is a non-zero constant with respect to $\sf{SNR}$.
Hence, a lower bound on the outage probability can be written as
\begin{equation}\label{outage probability upper bound}
\begin{split}
{\Pr} _{mmse}^{out}\left( \mathcal {R} \right) &\ge \Pr \left( {M \log \left( \frac{1 + \frac{{\sf{SNR}}}{{{\sf{INR}} }}{\lambda _M}}{{M}}{\sum\limits_{i =
1}^M\frac{1}{{|{{u_{M,i}}}|}^2}} \right) \le \mathcal {R}} \right)\\
 &\ge \Pr \left( \mathcal {A} \right)\Pr{\left( \log \left( {1 + \frac{{\sf{SNR}}}{{{\sf{INR}} }}{\lambda _M}} \cdot c \right) \le
\frac{\mathcal {R}}{M} \right)}\\
 &\doteq \Pr \left( 1 + {\frac{{\sf{SNR}}}{{{\sf{INR}}}} \cdot {\lambda _M} \cdot c \le {\sf{SNR}}^{\frac{{r}}{M}}}
\right).
\end{split}
\end{equation}
It could be immediately
verified that the last line of (\ref{outage probability upper bound}) and the last line of (\ref{asymptotic outage probability upper bound}) are asymptotically equivalent. Therefore, using the same argument as in (\ref{solution of pout}), we obtain the outage exponent upper bound as
\begin{equation}\label{outage exponent upper bound}
{{d}_{mmse}^{*}}\left( {r},\xi \right) \le \left( {N - M + 1}
\right)\left( {1 - \xi - \frac{{r}}{M}} \right)^{+}.
\end{equation}
Thus, the outage exponent upper bound equals to the lower bound.
This completes the proof.

\section{Discussion and Numerical results}

\begin{figure}[!tp]
\centering
$\!\!\!\!\!\!\!\!\!\!\!\!\!\!\!\!\!\!$ \includegraphics[width=4.5in]{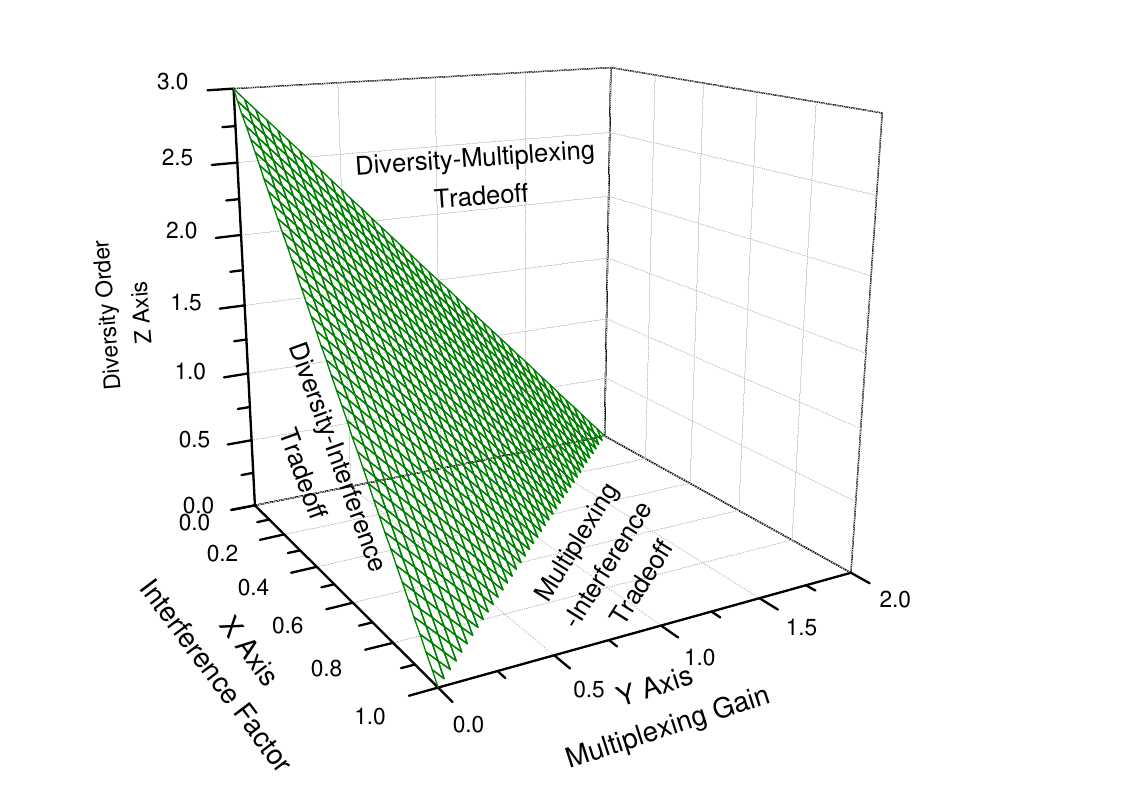}
\caption{Function ${{d}}^*\left( {{r}}, {\xi} \right)$ with $M$ = 2
and $N$ = 4.} \label{DMT function with interference}
\end{figure}

Theorem 1 in the previous section characterized the fundamental
tradeoff among the diversity order, multiplexing gain and inter-cell
interference factor. Given the DMT results in point-to-point
scenarios as \cite{DMT MMSE:Kumar}
\begin{equation}\label{DMT in point-to-point scenario}
{{d}_{mmse}^{*}}\left( {r} \right) = \left( {N - M + 1}
\right)\left( {1 - \frac{{r}}{M}} \right)^+,
\end{equation}
we can rewrite (\ref{MMSE DMT}) as
\[{{d}_{mmse}^{*}}\left( {r}, \xi \right) = \left( {N -
M + 1} \right)\left( {1 - \xi - \frac{{r}}{M}} \right)^+\]
\[ \quad\quad\quad\quad\:\,\,= {{d}_{mmse}^{*}}\left( {r} \right) - \left( {N - M +
1} \right)\xi \]
\begin{equation}\label{DMT analysis}
\: = {{d}_{mmse}^{*}}\left( {r} + M\xi \right).
\end{equation}
Obviously, the DMT performance in point-to-point scenarios can be
characterized as a special case of the DMT performance in cellular
systems when $\xi = 0$. For $\xi \neq 0$, according to (\ref{DMT
analysis}), we can either degrade the diversity order for $\left( {N
- M + 1} \right)\xi$ to maintain the multiplexing gain or pull down the
degrees of freedom for $M\xi$ but guarantee the diversity order. Actually, this is due to the fact that the introduced
interference reduces the minimum distance between the constellation
points with a scale factor related to ${\xi}$. Specifically, if $(N
- M + 1)\xi > M\xi$, i.e. ${N
> 2M - 1}$, the diversity order will suffer more loss from the
inter-cell interference than the multiplexing gain, and vise versa.
In Fig.~\ref{DMT function with interference}, we plot the function
${{d}}^*\left( {{r}}, {\xi} \right)$ for $M = 2$ and $N = 4$. The
slope of the intersection line between the result plane and the
$Z$-$X$ plane describes the rate of change of ${d}$ with respect to
${\xi}$. And the slope of the intersection line between the result
plane and the $Y$-$X$ plane describes the rate of change of ${r}$
with respect to ${\xi}$.

\begin{figure}[!tp]
\centering
$\!\!\!\!\!\!\!\!\!\!\!\!\!\!\!\!\!\!$\includegraphics[width=4.5in]{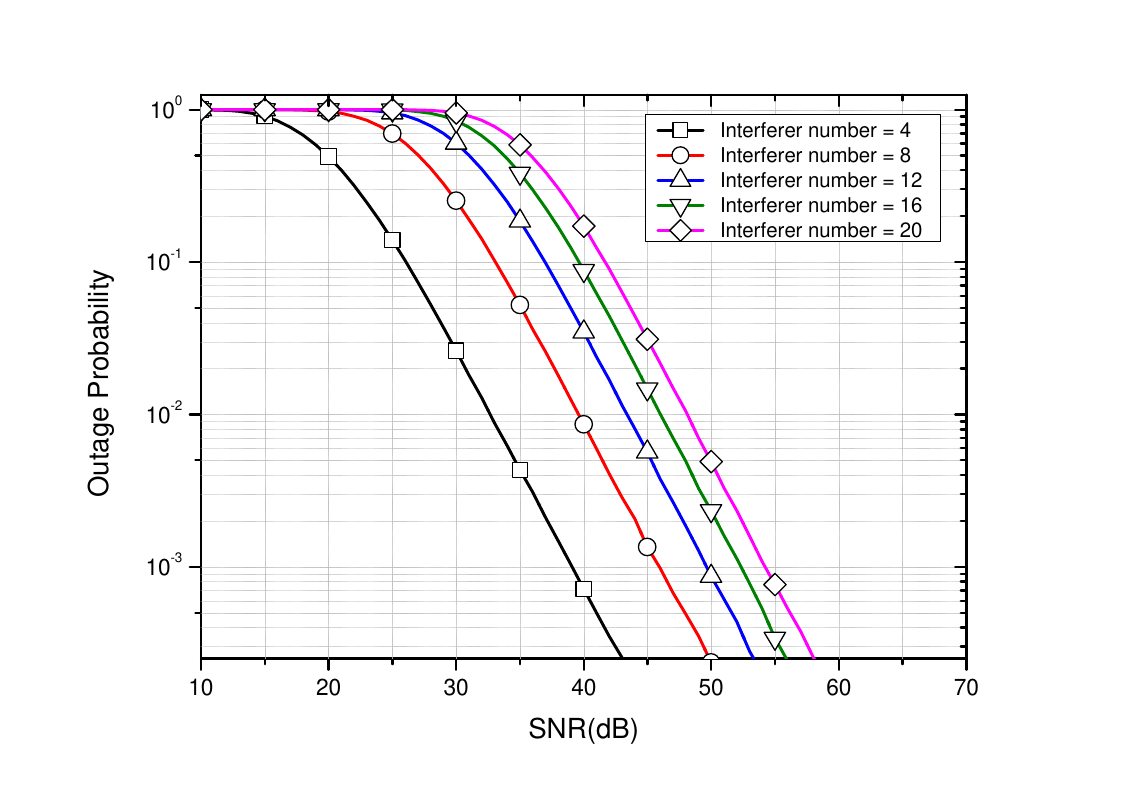}
\caption{Outage probability of MMSE receiver with different
interferer number}
\label{different interferer number}
\end{figure}

Another implication from Theorem 1 gives the fact that the DMT
performance has nothing to do with the number of the interferers. To verify this point, Monte Carlo simulation is employed and the corresponding results are shown in Fig.~\ref{different interferer number}, where $M = 2$, $N = 4$, $\xi = 0.5$ and the target rate
$\mathcal {R}=5$ bits per channel use. We see that the diversity
order keeps unchanged as the number of the interferers increases. And indeed, this observation implies that the DMT analysis cannot fully characterize the tradeoff among the outage probability, data rate and the interference. In Fig.~\ref{different interferer number}, though DMT remains the same for different number of interferers, for a given outage probability, a considerable increment on SNR is required to support the target rate when more interferers are invloved. Actually, according to \cite{DMT Comment:Angel Lozano}, DMT is only a coarse description of the fundamental tradeoff between the outage probability and the data rate. It asymptotically captures the rate of change of the outage probability and the data rate versus SNR in the high-power regime, but ignores the constant offset. Thus, though the number of the interferers has no effect on DMT, it affects the constant offset of the outage probability. And the exact impact of the number of the interferers on the system reliability and efficiency would be addressed in our future work.


\section{Conclusions}

In this paper, we study the diversity and multiplexing tradeoff in
the uplink of cellular systems to evaluate the impact of inter-cell
interference on the system reliability and efficiency. When suboptimal
linear MMSE receiver is utilized, fundamental
tradeoff among diversity order ${d}$, multiplexing gain ${r}$ and
inter-cell interference factor $\xi$ is asymptotically characterized
as ${d}_{{mmse}}^*\left( {r},\xi \right) = \left(N - M +
1\right)\left({1} - \xi - \frac{{r}}{M}\right)^{+}$. Given ${N
> 2M - 1}$, it is shown that the diversity order will suffer more loss from the inter-cell
interference than the multiplexing gain, and vise versa. We also observe the fact that, due to the weakness of the DMT characterization, the number of the interferers has no effect on DMT. Therefore, a more complete picture of the impact of the interference is needed. In summary,
in cellular MIMO systems, we should take both the randomness of the
channel and the inter-cell interference into consideration to
achieve the optimized data rate and error probability.

\section*{Acknowledgment}
This work was supported in part by the China 863 Program under Grant 2011AA100706, the
NSFC under Grants 61171107, 60972073 and 60871042, the China National Great Science Specific
Project under Grant 2010ZX03001-003.
%

\end{document}